\documentclass[aps,prb,twocolumn,floatfix,longbibliography,nofootinbib,superscriptaddress]{revtex4-1}
\usepackage[utf8]{inputenc}
\usepackage{natbib}
\usepackage{graphicx}
\usepackage{xcolor}
\usepackage[tbtags]{amsmath}
\usepackage[colorlinks, linkcolor=blue]{hyperref}
\hypersetup{colorlinks,allcolors=black}
\usepackage{amssymb}
\usepackage{amsmath}
\usepackage{gensymb}
\usepackage{float}
\usepackage{amsmath}
\usepackage{tabularx,graphicx}
\usepackage{epstopdf}
\usepackage{latexsym}
\usepackage{color, colortbl}
\usepackage{psfrag}
\usepackage{bbm}
\usepackage{bm}
\usepackage{lmodern}
\usepackage{fix-cm}
\usepackage{xfrac}
\usepackage{color}    
\usepackage{lipsum}
\usepackage{textcomp}
\usepackage{amsmath}
\usepackage{mathtools}
\usepackage{subfigure}
\usepackage{titlesec}
\usepackage{dsfont}
\usepackage{feynmp}
\usepackage{slashed}
\usepackage{multirow}
\usepackage[normalem]{ulem}

\newcommand{\bcen}{\begin{center}}
\newcommand{\ecen}{\end{center}}
\newcommand{\btab}{\begin{tabular}}
\newcommand{\etab}{\end{tabular}}
\newcommand{\bdes}{\begin{description}}
\newcommand{\edes}{\end{description}}

\newcommand{\beq}{\begin{equation}}
\newcommand{\eeq}{\end{equation}}
\newcommand{\bea}{\begin{eqnarray}}
\newcommand{\eea}{\end{eqnarray}}

\newcommand{\bary}{\begin{array}}
\newcommand{\eary}{\end{array}}
\newcommand{\benum}{\begin{enumerate}}
\newcommand{\eenum}{\end{enumerate}}
\newcommand{\bitem}{\begin{itemize}}
\newcommand{\eitem}{\end{itemize}}

\newcommand{\eqn}[1] {eqn.~(\ref{#1})}

\newcommand{\Fig}[1]{Fig.~\ref{#1}}

\makeatletter

\newcommand{\Rmnum}[1]{\expandafter\@slowromancap\romannumeral #1@}
\makeatother

\begin{document}

\title{Classical and quantum facilitated exclusion processes}
\author{Amit Kumar Chatterjee}
\affiliation{Yukawa Institute for Theoretical Physics, Kyoto University,  Kyoto 606-8502, Japan}
\author{Adhip Agarwala}
\affiliation{Indian Institute of Technology Kanpur, Kalyanpur, Kanpur, India 208 016}

\begin{abstract}
We demonstrate exciting similarities between classical and quantum many body systems whose microscopic dynamics are composed of non-reciprocal three-site facilitated exclusion processes. We show that the quantum analog of the classical facilitated process engineers an interesting {\it quantum absorbing transition} where the quantum particles transit from an unentangled direct-product absorbing phase to an entangled steady state with a finite current at density $\rho=1/2$. In the generalized classical facilitated exclusion process, which includes independent hopping of particles with rate $p$, our analytical and Monte-Carlo results establish emergence of a special density $\rho^*=1/3$ that demarcates two regimes in the steady state, based on the competition between two current carrying modes (facilitated and independent). The corresponding quantum system also displays similar qualitative behaviors with striking non-monotonic features in the bipartite entanglement. Our work ties the two sub-fields of classically interacting exclusion processes, and interacting non-Hermitian quantum Hamiltonians to show common themes in the non-equilibrium phases they realise.

\end{abstract}
 
\maketitle

{\it Introduction:} Classical and quantum non-equilibrium phenomena has been a cornerstone of all of physics. In the classical domain, their applicability has ranged from understanding transport phenomena \cite{Schadschneider_2011}, out-of-equilibrium materials 
\cite{Giannetti_2016,Tena_Solsona_2017}, disease-modelling \cite{Kamenev_2008}, game theory \cite{Babajanyan_2020}, biological systems \cite{Allman_2003,Blythe_2007} and 
even socio-economic structures \cite{Castellano_2009}. In the quantum regime, such systems have been studied in context of open quantum  systems (such as cold-atoms\cite{MaciejAOP2007,BlochRMP2008,LangenARCMP2015,LiangPRL2022}), and have been recently explored in random quantum circuits \cite{Fleckenstein_PRR_2022}  and broadly in context of non-Hermitian physics \cite{Lieu_PRA-2019,LieuPRB19, Panda20, Carlstrom20}. While the classical and quantum community has been working on the various aspects of such systems, their motivations and quantities of interest have also been varied. For instance, while in classical Markovian systems, non-equilibrium steady states \cite{Blythe_2007_mpa,Zhang_2012,Ge_2012} as well as relaxations towards steady states \cite{Krapivsky_2010,Evans_2020,Lu_2017} have been centre of interests for interacting many particle systems; in quantum non-Hermitian phenomena largely the focus has been on single-particle spectra and their topological properties \cite{alvarez2018topological,ghatak2019new,torres2019perspective,ashida2020non,bergholtz2021exceptional, Gong18, Kawabata19,lee2019anatomy}. Very recently interacting non-Hermitian Hamiltonians \cite{Kawabata_PRB_2022, Kawabata_PRL_2021, Fukui_PRB_1998, Hatano_PRL_1996, Zuo_PRB_2022, Crippa_PRB_2021, Xu_PRB_2020, Panda20, Liu_PRB_2020, Mu_PRB_2020, Yamamoto_PRL_2019, Nakagawa_PRL_2018, Zhang_PRB_2022, BanerjeePRB22022, Zhang_PRB_2021, Yang_PRB_2022, Pan_PRA_2019, Solofo_PRR_2021, Ghosh_PRB_2022, HeussenPRB2021}, and the long-time steady states  they may realize have been studied -- however the field remains broadly unexplored. 

Investigations on the common themes between the physics of such classical-quantum problems have been far and few \cite{Horsenn_PRB_2015, Kavanagh_2022,Hruza_arXiv_2022,berthier2019zero}, for instance only recently the rare trajectories of a Markovian system was mapped to the ground states of a quantum topological Hamiltonian\cite{Garrahan_arXiv_2022}. This broadly builds on idea that Markov matrices of the classical rate processes could be mapped to a sub-class of constrained quantum many body Hamiltonians. But apart from deriving such structural equivalence between the mathematical structures not a lot of work has gone into finding common physics in the non-equilibrium phases they realize.

\begin{figure}[h!]
    \centering
 \subfigure[]{\includegraphics[width=0.85\columnwidth]{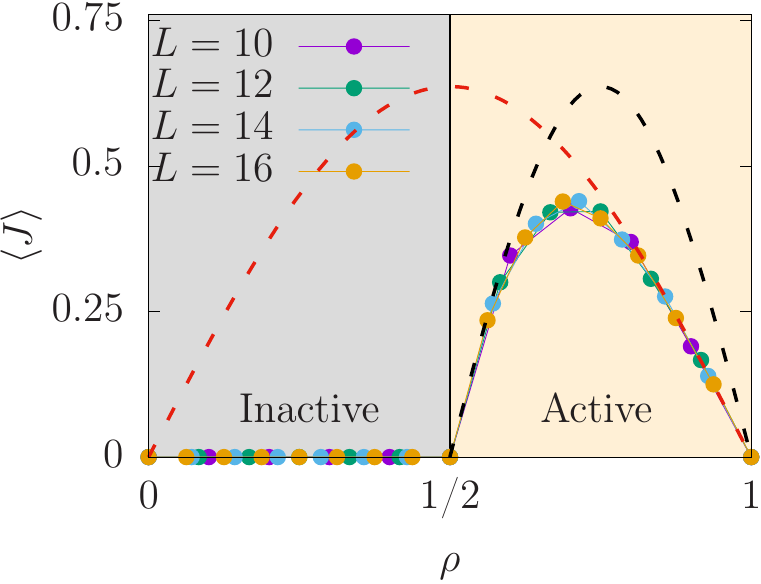}}
  \subfigure[]{\includegraphics[width=0.85\columnwidth]{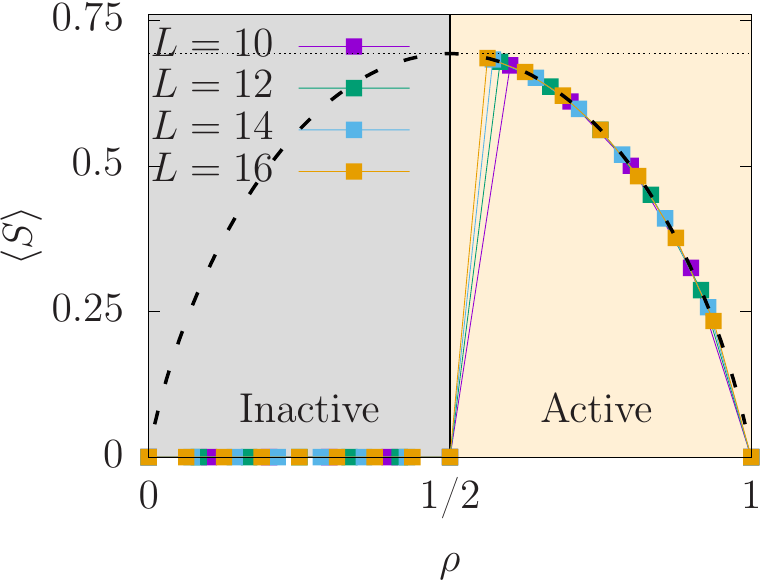}}
    \caption{ {\bf  QF-TASEP:} (top) The variation of steady state current as a function of density of particles $\rho$ in the quantum version of F-TASEP (see \eqn{Qrasep}). The dashed lines corresponds to mean-field expectations (see text). (bottom) The bipartite entanglement entropy of the first site with the rest of the system in the steady state as a function of density. Dashed line corresponds to the result for a free Fermi gas.}
    \label{fig:RASEP}
\end{figure}

In this work, we demonstrate intriguing resemblance in macroscopic physical characteristics between classical and quantum many body systems, whose  dynamics are based on a three site microscopic process in one dimension. This three site process, familiar as {\it facilitated totally asymmetric simple exclusion process} (F-TASEP), allows unidirectional hopping of particle to a vacant neighbor only if its other neighbor is occupied. The F-TASEP,  interestingly, is known to exhibit absorbing phase transition between active and inactive phases with $\rho=\frac{1}{2}$ being the transition point where $\rho$ is the conserved particle density. We introduce QF-TASEP, the quantum version of F-TASEP, for which the bipartite entanglement and current remarkably exhibit quantum absorbing phase transition at $\rho=\frac{1}{2}$ with active and inactive phases, similar to the classical F-TASEP. Such active-inactive phase transitions disappear for both classical and quantum cases as we generalize the F-TASEP by including the independent hopping (i.e. not facilitated by neighboring particles) of particles with rate $p$, consequently termed as $p$F-TASEP (classical) and $p$QF-TASEP (quantum) respectively.  Due to the presence of multiple current carrying modes ({\it facilitated} and {\it independent}), we compare their individual contributions to the total steady state particle current in the $\rho-p$ plane. Interestingly, our exact analytical calculations for $p$F-TASEP reveals the emergence of a special point $\rho=\frac{1}{3}$ that demarcates between two different regimes. For the system density with any fixed value between $0<\rho\leqslant\frac{1}{3}$, the independent current always dominates over the facilitated current for any $p$. On the contrary, for $\frac{1}{3}<\rho<1$, there is always a finite $p^\ast$ such that the facilitated (independent) current dominates for $p<p^\ast$ ($p>p^\ast$). Remarkably, in the quantum problem $p$QF-TASEP, the bipartite entanglement shows striking features around $\rho=\frac{1}{3}$ indicating towards its speciality similar to the classical case.

{\it Model:} We start with the toy model F-TASEP  defined on a one dimensional periodic lattice with the following microscopic dynamics 
\begin{equation}
110 \,\,\stackrel{1}{\longrightarrow}\,\,  101,
 \label{eq:rasep}
\end{equation}
where $1$ and $0$ denote particle and vacancy respectively and each lattice site can be occupied by at most one particle. The model is a special instance of the Katz-Lebowitz-Spohn model \cite{Katz_1984,Hager_2001} broadly used to model superionic conductors \cite{Dieterich_1980,Katz_1984}, vehicular traffic flow \cite{Gray_2001,Levine_2004} etc. Notably, alongside non-equilibrium physics literature \cite{Basu_PRE_2009,Gabel_2010,Antal_2000,Shi_2012,Hao_2016,Chatterjee_2018,Botto_2018,Pelizzola_2019}, the F-TASEP and its variations have gained lot of recent interests in the area of probability and mathematics \cite{Baik_2018,Goldstein_2019,Goldstein_2021,Blondel_2020,Blondel_2021,Goldstein_2022,Ayyer_2020,Barraquand_2023}. The F-TASEP as discussed in \cite{Basu_PRE_2009}, shows a finite density transition at $\rho_c=\frac{1}{2}$ where the system transits from inactive and non-ergodic state to a current carrying and ergodic steady state. This active-inactive transition is characterised by the order parameter $\langle 110 \rangle$ which actually measures the activity (density of active sites) or current in the system. The order parameter remains zero for $0\leqslant\rho\leqslant\frac{1}2$ whereas it is non-zero for $\rho>\frac{1}{2}$ and it varies as $(\rho-\rho_c)^\beta$ with the exponent $\beta=1$.  

We now study the corresponding quantum process QF-TASEP by introducing the following non-Hermitian quantum Hamiltonian given by
\beq
H = \sum_i n_{i-1} c^\dagger_{i+1} c_{i}
\label{Qrasep}
\eeq
Given a set of many body eigenvalues $E_n$, the quantum steady state is given by the eigenstate whose imaginary component of the eigenvalue is maximum such that given the time evolution of every many body state $|\psi_n\rangle$ with energy $E_n$ the state grows as a function of time ($|\psi_n(t)\rangle = \exp(-i E_n t) |\psi_n\rangle$) and gets chosen as the steady state at infinite time.

{\it Quantum Absorbing Phase Transition:} Investigating the steady states of QF-TASEP (see \eqn{Qrasep}) and measuring the current operator  
\begin{equation}
J_Q=\frac{1}{L}\sum_j -i \langle c^\dagger_{j+1} c_j - c^\dagger_j c_{j+1} \rangle 
\label{eq:jq}
\end{equation}
on the non-equilibrium quantum steady state shows a similar absorbing transition as a function of density -- where the system transits to a current carrying steady state when $\rho > \frac{1}{2}$ (see \Fig{fig:RASEP}). Remarkably, although the quantum current operator defined in Eq.~(\ref{eq:jq}) is different from the usual classical current $J=\langle 110 \rangle$ for F-TASEP, they essentially capture similar characteristics.  For $\rho<\frac{1}{2}$, the correlated hopping process as described above, given any initial state forms charge density wave patches where every time a dipole is formed (01), the region is rendered inactive. This charge-density wave patch spans the complete lattice at $\rho=\frac{1}{2}$ -- at which any additional particle leads to a finite current steady state. This active-inactive transition can be also found in the density of active sites $\langle  n_{i-1} n_i (1-n_{i+1}) \rangle$, which is basically the classical order parameter. Given the quantum nature of the state, it would be interesting to ask if we can find the signature of quantum absorbing phase transition in a bipartite entanglement. Indeed, it shows a discontinuous jump at $\rho=\frac{1}{2}$ as shown in \Fig{fig:RASEP}. A trivial mean-field where the $n_i \rightarrow \rho$ (average density) reduces to an effective Hatano-Nelson model \cite{Hatano_PRL_1996} where the hopping is renormalized by density. This leads to a boosted Fermi sea steady state where the finite momentum modes lead to a finite current state  with $J_Q = \frac{2}{\pi} \sin(\rho \pi)$, as seen in the dashed line in \Fig{fig:RASEP}(a) where the current matches the analytical result quite well. Near $\rho \sim 0.5$, we have effective carriers as $\tilde{\rho}=\rho-\frac{1}{2}$ which matches the current profile near $\rho \sim \frac{1}{2}$ where $J_Q \sim \frac{2}{\pi}\sin(2\tilde{\rho}\pi)$. It is interesting that this dynamical transition has no symmetry broken order parameter but current itself can be thought of as an order parameter, with the critical exponent $\beta=1$ [$J_Q \sim (\rho-\rho_c)^\beta$] same as that of the classical case. Interestingly, the bipartite entanglement entropy of a site with rest of the system, follows $S(\rho) = - \Big( \rho \log(\rho) + (1-\rho) \log(1-\rho) \Big)$ (similar to a free Fermi gas)
(see the analytical curve in \Fig{fig:RASEP}(b)) as soon as $\rho>\frac{1}{2}$, showing a discrete jump in the entanglement entropy characteristic of this quantum transition.

\begin{figure}
    \centering
    \includegraphics[width=0.8\columnwidth, angle=90]{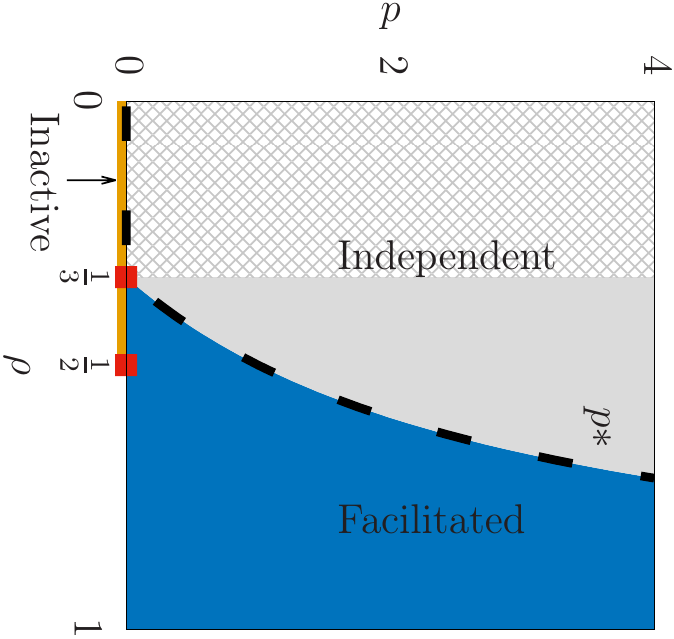}
    \caption{{\bf $p$F-TASEP phase diagram:} Analytically obtained phase diagram for the $p$F-TASEP (see \eqn{eq:fai-tasep}) with both facilitated and independent hopping. The figure compares the steady state contributions of the two different current modes $J_{fa}$ and $J_{in}$. The point $\rho=\frac{1}{3}$ distinguishes between two different regimes. For $0<\rho\leqslant\frac{1}{3},$ the independent current is always dominant irrespective of the value of $p$. In sharp contrast, for $\frac{1}{3}<\rho<1$, there exists a finite $p^\ast(\rho)$ (see \eqn{eq:phaseline}) that decides which mode (facilitated or independent) will be dominant for a fixed density.}
    \label{fig:phase_dia_ftasep}
\end{figure}

{\it Generalization to $p$F-TASEP.-} Both the classical F-TASEP and quantum QF-TASEP exhibit absorbing phase transition which for $\rho\leqslant\frac{1}{2}$, is identified by the non-ergodicity in the system due to the presence of multiple absorbing or inactive configurations. However, such absorbing phase transition vanishes as we generalize the dynamics of F-TASEP [Eq.~(\ref{eq:rasep})] by initiating another mode of hopping, namely the {\it independent} hopping of particles which occurs without any facilitation from the neighbor. This generalized classical process, which we call $p$F-TASEP, is defined as 
\begin{equation}
110 \,\,\stackrel{1}{\longrightarrow}\,\,  101, \hspace*{1 cm} 010 \,\,\stackrel{p}{\longrightarrow}\,\,  001.
 \label{eq:fai-tasep}
\end{equation}  
Clearly, the limits $p=0$ and $p=1$ of $p$F-TASEP correspond to F-TASEP and TASEP respectively. The independent hopping of particles with rate $p$ restores ergodicity in the system for any density.  The model has been studied previously as a cooperative exclusion process, in context of time evolution of a step initial condition leading to rich phenomena like formation of shocks and compression or rarefaction waves \cite{Gabel_2011}. Our focus for $p$F-TASEP would be to make a comparative study of the individual contributions to steady state current produced by two different modes, {\it facilitated current} $J_{fa}$ and {\it independent current} $J_{in}$.  Analysis of multiple current carrying modes become naturally important for wide range of scenarios e.g. multiple chemical reagents diffusing through narrow channels and undergoing specific reactions among selective pairs, different types of vehicles moving through same road, decision making in teams with sub-groups etc. We reformulate the Ising model like pair-factorized steady state of $p$F-TASEP as a non-equilibrium matrix product state by obtaining the particles and vacancies as two non-commuting matrices obeying certain matrix algebra (see \cite{supp} for details). These matrix algebra help us to analytically calculate the observables of interest, $J_{fa}=\langle 110\rangle$ and $J_{in}=p\langle 010\rangle$. The exact expressions are obtained to be
\begin{eqnarray}
J_{fa}(\rho,p)&=&\frac{z(2\rho-1)}{(1-p)(z-p)}\frac{z(1-\rho)-p\rho}{z\rho-p(1-\rho)},\cr\cr
J_{in}(\rho,p)&=&\frac{p(1-\rho)}{(1-p)}\frac{z(1-\rho)-p\rho}{z\rho-p(1-\rho)},\cr
\mathrm{where} \hspace*{0.2 cm} z(\rho,p)&=& \frac{1}{2\rho(1-\rho)}[1+2\rho(1-\rho)(p-2)\cr && -(1-2\rho)\sqrt{1-4\rho(1-\rho)(1-p)}].
\label{eq:jfajin} 
\end{eqnarray}
To extract the competition between the two current carrying modes in $\rho-p$ plane, one has to simply equate $J_{fa}$ and $J_{in}$ from Eq.~(\ref{eq:jfajin}) and consequently find the curve that separates the region of dominance between the facilitated mode and the independent mode in the $\rho-p$ plane. This leads us to the following non-trivial solution 
\begin{equation}
p^\ast(\rho)=\frac{3\rho-1}{1-\rho}.
\label{eq:phaseline}
\end{equation}
\begin{figure}[t]
  \centering
  \subfigure[]{\includegraphics[width=0.85\columnwidth]{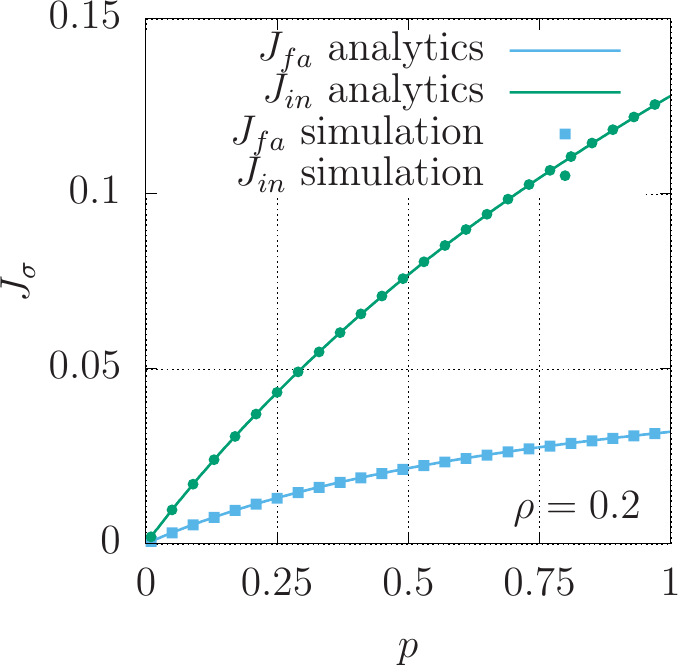}}
  \subfigure[]{\includegraphics[width=0.85\columnwidth]{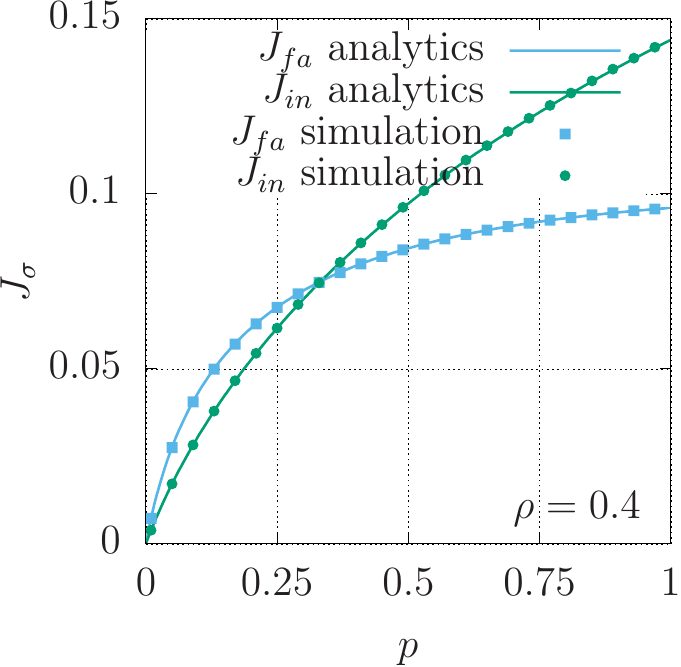}}
  \caption{ {\bf Facilitated and Independent steady state currents:} The  behavior of facilitated steady state current ($J_{fa}$) and independent current ($J_{in}$) as a function of $p$ at $\rho = 0.2$ ($<\frac{1}{3}$) in (a) and for $\rho= 0.4$ ($>\frac{1}{3}$) in (b). Analytical results are shown with a line and Monte Carlo results (ensemble averages are  performed over $10^6$ configurations) are shown in dots.  For $\rho>1/3$ there is a finite $p$ below which $J_{fa} >J_{in}$.}
\label{fig:cutplots}
\end{figure}
The interesting fact evident from Eq.~(\ref{eq:phaseline}) is that no physical solution for $p^\ast(\rho)$ exists for $\rho\leqslant\frac{1}{3}$. This implies that, for $p$F-TASEP with conserved density $0\leqslant\rho\leqslant\frac{1}{3}$, one mode has always higher contribution to current than the other irrespective of how small or large $p$ is. We identify this dominant mode as the independent mode in Fig.~\ref{fig:phase_dia_ftasep} where we present the comparison between two current carrying modes in the $\rho-p$ plane. Contrary to $0\leqslant\rho\leqslant\frac{1}{3}$, for any conserved system density $\frac{1}{3}\leqslant\rho\leqslant1$, there is a finite $p^\ast$ such that $J_{fa}$ dominates over $J_{in}$ for $p<p^\ast$ whereas $J_{in}$ acts as the higher current carrying mode for  $p>p^\ast$. Clearly, as shown in Fig.~\ref{fig:phase_dia_ftasep}, $\rho=\frac{1}{3}$ emerges as a special density which demarcates between two regimes-- one with $p^\ast=0$ (for $0\leqslant\rho\leqslant\frac{1}{3}$) and the other with $p^\ast>0$ (for $\frac{1}{3}\leqslant\rho\leqslant1$). Note that the demarcator point $\rho=\frac{1}{3}$ for $p$F-TASEP is different from the absorbing transition point $\rho=\frac{1}{2}$ for F-TASEP and QF-TASEP. 

To show the individual behaviors of $J_{fa}$ and $J_{in}$ explicitly, as functions of the parameter $p$, we present Fig.~\ref{fig:cutplots} (a)-(b). In both figures, we observe that our exact analytical results (solid lines) agree with Monte Carlo simulation results (dots). The Fig.~\ref{fig:cutplots}(a) shows that the independent current is always higher than the facilitated current as expected for  $\rho<\frac{1}{3}$. In contrast, we observe in Fig.~\ref{fig:cutplots}(b) ($\rho>\frac{1}{3}$)  that $J_{fa}$ and $J_{in}$ cross each other at $p=p^\ast$. 

\begin{figure}
    \centering
  \subfigure[]{\includegraphics[width=0.85\columnwidth]{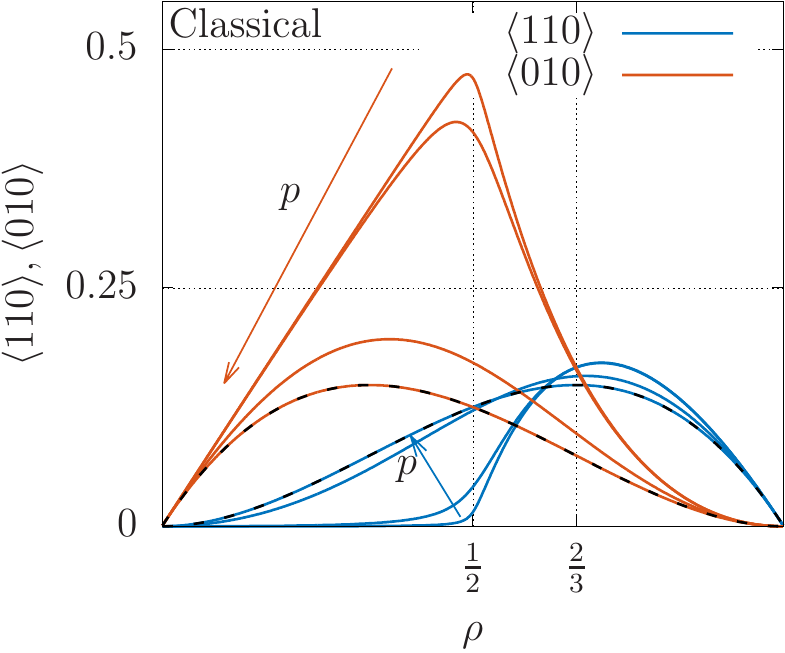}}
  \subfigure[]{\includegraphics[width=0.85\columnwidth]{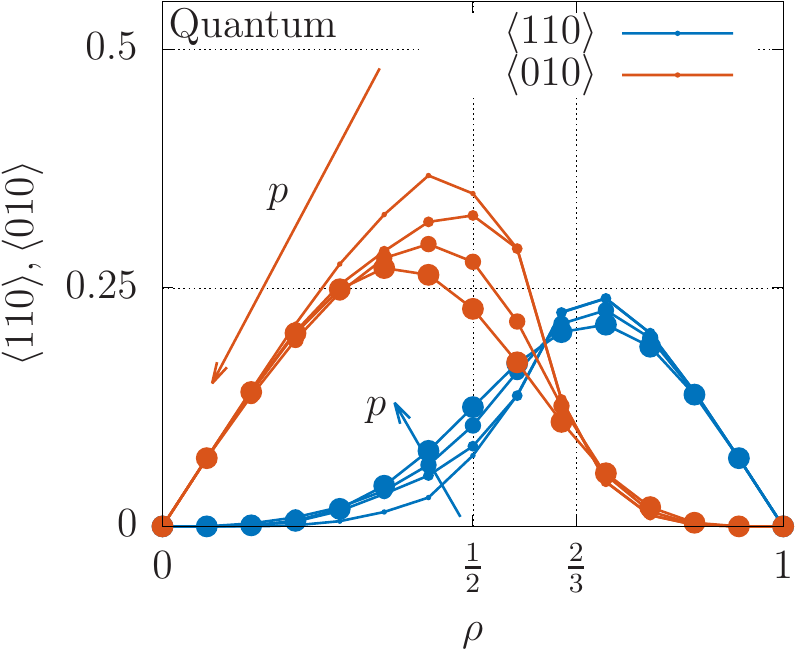}}
    \caption{{\bf $p$F-TASEP and $p$QF-TASEP}: Comparison between the facilitated activity and independent activity in the classical and quantum facilitated processes for different values of $p$ ($0.001,0.01,0.5,1.0$). (Analytical results from MPS in classical and numerical ED for $L=14$ in the Quantum case)}
    \label{fig:compare}
\end{figure}

{\it  Quantum analogue of $p$F-TASEP.-} In order to develop $p$QF-TASEP, the quantum process corresponding to classical $p$F-TASEP, we study the following non-Hermitian Hamiltonian 
\begin{equation}
 H = \sum_i n_{i-1} c^\dagger_{i+1} c_{i} + p (1-n_{i-1}) c^\dagger_{i+1} c_{i}.
 \label{pftasep}
\end{equation}
As observables, we analyze $\langle 110\rangle$ and $\langle 010\rangle$ which are directly related to the facilitated and independent currents that have played crucial roles in the discussions in the previous section. In connection to the activity $\langle 110 \rangle$ for F-TASEP and QF-TASEP, we can term $\langle 110 \rangle$ and $\langle 010 \rangle$ as {\it facilitated activity} and {\it independent activity}, respectively,  in cases of $p$F-TASEP and $p$QF-TASEP.  For these expectation values, we show the analogy between the classical and quantum cases in \Fig{fig:compare}. Since at $p=1$ the $p$F-TASEP reduces to the usual TASEP that has uncorrelated steady state, classically both these expectation values behave as $\rho^2(1-\rho)$ and as $(1-\rho)^2 \rho$ as shown via the dashed curves in \Fig{fig:compare} (top). 
With decreasing $p$, the independent hopping decreases, thereby increasing $\langle 010\rangle$ for some fixed density. On the other hand, $\langle 110\rangle$ decreases with decreasing $p$ since the production rate of nearest neighbor particle pair $\langle 11 \rangle$ through the process $0101\stackrel{p}{\longrightarrow}0011$ reduces. The situation becomes most dramatic at $p\rightarrow 0$ where $\langle 110\rangle$ shows the near-critical behavior reflective of the absorbing phase transition. Interestingly, similar qualitative behaviors are observed for $p$QF-TASEP (see \Fig{fig:compare} (bottom)), where we study the expectation value of the two operators $\langle n_i n_{i+1} (1-n_{i+2}) \rangle$ and $\langle (1 - n_i) n_{i+1} (1-n_{i+2}) \rangle$.  Thereby, the quantum steady state fascinatingly shows a behavior equivalent to the classical non-equilibrium steady state.

\begin{figure}
    \centering
    \includegraphics[width=0.9\columnwidth]{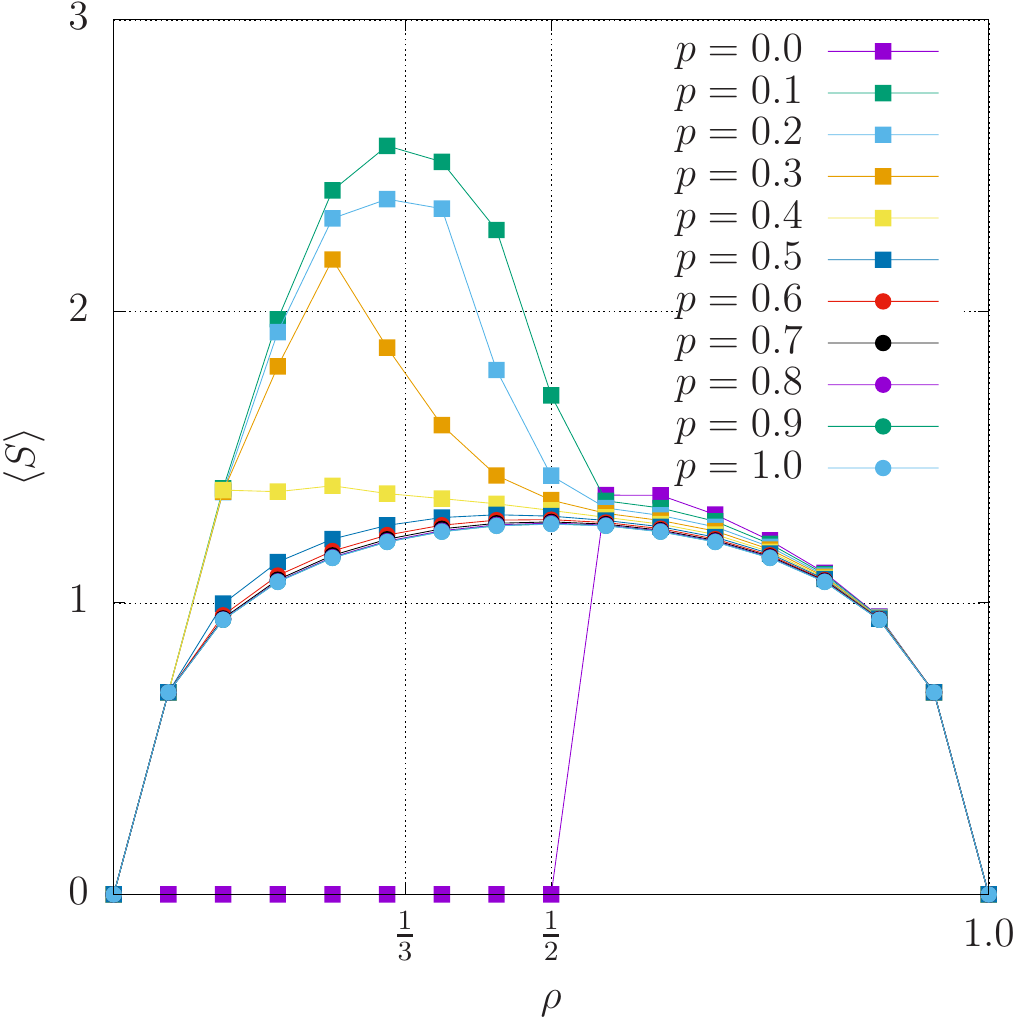}
    \caption{{\bf Enhanced bipartite entanglement:} The behavior of the bipartite entanglement entropy of half of the system with the other half, as a function of density $\rho$ for the pQF-TASEP (see \eqn{pftasep}) for different values of $p$ for $L=16$.}
    \label{entan}
\end{figure}

We next investigate the bipartite entanglement entropy of this system. The corresponding behavior is shown in \Fig{entan}. Interestingly at $p=1$ when the system is just the quantum Hatano-Nelson model, the entanglement behaves characteristic of a equilibrium Fermi sea where $S(\rho) \sim \log(L \rho(1-\rho))$ since the steady state can be thought of as the boosted Fermi sea. Similarly at $p=0$ when the system shows a quantum active-inactive transition at $\rho \sim \frac{1}{2}$ one finds that $\rho < \frac{1}{2}$, $\langle S \rangle $=$0$ as is expected for an absorbing state. What is interesting is the rise in entanglement at small values of $p$ for less than half filling in the system. We suspect that this rise is due to the formation of doublons in the system. For instance at $p\rightarrow 0$, the doublon state $|110\rangle$ disperses with a second order process with an effective non-Hermitian Hamiltonian $\sim p e^{ik}$. Notably, the entanglements for sufficiently small values of $p$ exhibit maximums near $\rho=\frac{1}{3}$. The striking behaviors of the bipartite entanglement entropy for $p$QF-TASEP and the relation between $p$F-TASEP and $p$QF-TASEP in context of $\rho=\frac{1}{3}$ being a special point, will be explored further in the future work.

{\it Conclusion.-}In this work we study similarities in macroscopic behaviors of classical and quantum many body systems, based on a three site microscopic process, the facilitated exclusion process. Remarkably, analogous to the absorbing phase transition shown by classical F-TASEP, we find that the quantum facilitated process Q-FTASEP exhibits quantum absorbing phase transition characterized by the bipartite entanglement entropy and current. Notably, the absorbing transition for both the classical and quantum models occur at $\rho=\frac{1}{2}$ and the current, acting as the order parameter, varies as $(\rho-\rho_c)^\beta$ with $\beta=1$ for both cases. The absorbing transition vanishes as we generalize the classical F-TASEP to $p$F-TASEP by including the independent hopping of particles with rate $p$.  Intriguingly, the comparison between the two sources of current (facilitated and independent)  reveals the existence of a special density $\rho=\frac{1}{3}$ which demarcates between two regimes, one where the independent current always dominates irrespective of $p$ value and the other where a finite $p$ distinguishes between the region of dominance of the two currents. The corresponding quantum model $p$QF-TASEP displays similar qualitative behaviors as the classical one and particularly the bipartite entanglement manifests striking features including sudden rise and maximal values near $\rho=\frac{1}{3}$. It would be interesting to further explore these classical-quantum connections for deeper understandings. Some immediate future directions include the investigations of classical-quantum connections with reciprocal hopping, open boundary conditions and relaxation phenomena. We believe that our present study paves paths to explore interesting connections for non-equilibrium phenomena overlapping between generic many-body classical systems and interacting non-Hermitian quantum Hamiltonians.

{\it Acknowledgements.-}  We thank Diptarka Das, Subrata Pacchal, Abhisodh Prakash, Arghya Das, Urna Basu, Hisao Hayakawa, Diptiman Sen for discussions. AA acknowledges support from IITK Initiation Grant (IITK/PHY/2022010) and workstation {\it Wigner} at IITK. A.K.C. gratefully acknowledges postdoctoral fellowship from the YITP and partial support from Grants-in-Aid for Scientific Research (Grant No. JP21H01006). We acknowledge use of open-source QuSpin\cite{weinberg2017quspin,weinberg2019quspin} for exact diagonalisation calculations. Several numerical calculations and simulations have been done on the cluster Yukawa-21 at YITP. 
Both of us acknowledge hospitality at ICTS, Bangalore. 

\bibliography{three_site_interaction.bib}

\newpage
\clearpage

\begin{widetext}

\begin{centering}
{\bf Supplemental Material for ``Classical and quantum facilitated exclusion processes"}

\vspace{1cm}

\end{centering}

In this supplementary material, we briefly describe the matrix product steady state of $p$F-TASEP, defined in the main text as
\begin{equation}
110 \,\,\stackrel{1}{\longrightarrow}\,\,  101, \hspace*{1 cm} 010 \,\,\stackrel{p}{\longrightarrow}\,\,  001.
\label{eq:s1} 
\end{equation}
The model obeying the microscopic dynamics Eq.~(\ref{eq:s1}) evolves according to the Master equation $\frac{\mathrm{d}|P\rangle}{\mathrm{d}t}=M|P\rangle$ where the vector $|P\rangle$ includes probabilities of all possible configurations $P(\left\lbrace s_1,\dots,s_L\right\rbrace)$. The Markovian rate matrix $M$ can be decomposed as
\begin{equation}
M=\sum_{i=1}^{L} I\otimes \dots I\otimes h_{i-1,i,i+1}\otimes I \dots \otimes I,
 \label{eq:s2}
\end{equation}
where $h_{i-1,i,i+1}$ is the local three site matrix with dimension $8\times8$ and $I$ is $2\times2$ identity matrix sitting at every site except the triad $(i-1,i,i+1)$. The off-diagonal elements of $h_{i-1,i,i+1}$ contain the transition rates between two different local three site configurations, whereas the diagonal terms carry the total outward rates from a three site configuration. The explicit form of the matrix $h_{i-1,i,i+1}$ for dynamics in Eq.~(\ref{eq:s1}) is
\begin{equation}
 h_{i-1,i,i+1}=\left( \begin{array}{cccccccc}
     0 & 0 & 0 & 0 & 0 & 0 & 0 & 0 \\
     0 & -1 & 0 & 0 & 0 & 0 & 0 & 0 \\
     0 & 1 & 0 & 0 & 0 & 0 & 0 & 0 \\
     0 & 0 & 0 & 0 & 0 & 0 & 0 & 0 \\
     0 & 0 & 0 & 0 & 0 & 0 & 0 & 0 \\
     0 & 0 & 0 & 0 & 0 & -p & 0 & 0 \\
     0 & 0 & 0 & 0 & 0 & p & 0 & 0 \\
     0 & 0 & 0 & 0 & 0 & 0 & 0 & 0 \\
    \end{array}
\right). \label{eq:sh}
\end{equation}
The steady state, by definition, implies $M|P\rangle=0$. We make the following matrix product ansatz, assuming that any configuration in the steady state can be represented as a product of matrices and the corresponding probability would be determined as
\begin{eqnarray}
P\left(\left\lbrace s_i \right\rbrace\right) &\propto& Tr\left[X_i\right], \cr
X_i&=& D \delta_{s_i,1} + E \delta_{s_i,0}, 
 \label{eq:s3}
\end{eqnarray}
with $D$ and $E$ representing particle and vacancy respectively. A sufficient condition to ensure the steady state is to consider the following local (probability) flux cancellation scheme
\begin{eqnarray}
&& h_{i-1,i,i+1} \left(\begin{array}{c}
                     D \\ E \\
                    \end{array}
\right)\otimes \left(\begin{array}{c}
                     D \\ E \\
                    \end{array}
\right)\otimes \left(\begin{array}{c}
                     D \\ E \\
                    \end{array}
\right) \cr&=&\left(\begin{array}{c}
                     D \\ E \\
                    \end{array}
\right)\otimes \left(\begin{array}{c}
                     \tilde{D} \\ \tilde{E} \\
                    \end{array}
\right)\otimes \left(\begin{array}{c}
                     D \\ E \\
                    \end{array}
\right)-\left(\begin{array}{c}
                     D \\ E \\
                    \end{array}
\right)\otimes \left(\begin{array}{c}
                     D \\ E \\
                    \end{array}
\right)\otimes \left(\begin{array}{c}
                     \tilde{D} \\ \tilde{E} \\
                    \end{array}
\right),\nonumber \\
 \label{eq:s4}
\end{eqnarray}
where $\tilde{D}$ and $\tilde{E}$ are known as {\it auxiliary} matrices that have to be chosen suitably such that the correct steady state is achieved. Such a suitable pair of choices for the present model turns out to be $\tilde{D}=-1$ and $\tilde{E}=0$. Along with these choices, from Eq.~(\ref{eq:s4}), we find the matrix algebra to be satisfied the matrices $D$ and $E$, are
\begin{eqnarray}
DDE&=&DE,\cr
pEDE&=&EE.
\label{eq:s5} 
\end{eqnarray}
A possible set of explicit matrix representations satisfying Eq.~(\ref{eq:s5}) are obtained to be
\begin{equation}
 \begin{array}{ccc}
  D=\left(\begin{array}{cc}
           0 & 1 \\
           0 & 1 \\
          \end{array}\right), & \hspace{1 cm} E=\left(\begin{array}{cc}
           p & 0 \\
           1 & 0 \\
          \end{array}\right).
 \end{array} \label{eq:s6}
\end{equation}
If we put $p=0$ (i.e. the F-TASEP) in the matrix algebra in Eq.~(\ref{eq:s5}) and  matrix representations in Eq.~(\ref{eq:s6}), the matrix algebra ($D^2=D$ and $E^2=0$) and matrix representations for the active phase in F-TASEP are correctly recovered \cite{Basu_PRE_2009}. With the explicit steady state matrix representations being obtained [Eq.~(\ref{eq:s6})], we can proceed to calculate observables, in particular the currents of the facilitated mode and independent mode,  $J_{fa}$ and $J_{in}$ respectively. These quantities can be calculated analytically as
\begin{eqnarray}
J_{fa}&=&\langle 110\rangle=\frac{Tr[DDET^{L-3}]}{Tr[T^L]},\cr
J_{in}&=&p\langle 010\rangle=p\frac{Tr[EDET^{L-3}]}{Tr[T^L]},\
\label{eq:s7}
\end{eqnarray}
where $T$ is the transfer matrix, translating which over the lattice sites generates all possible configurations, so that the partition function becomes $Tr[T^L]$. The matrix $T$ is defined as
\begin{equation}
T=zD+E=\left(\begin{array}{cc}
              p & z \\
              1 & z\\
             \end{array}
\right),
 \label{eq:s8}   
\end{equation}
where $z$ acts as the fugacity associated with the particles. As usual, the transfer matrix can be brought into diagonal form by $T_{d}=U^{-1}TU$ where the unitary matrix $U$ has eigenvectors of $T$ as its columns
\begin{equation}
U=\left(\begin{array}{cc}
         \lambda_--z & \lambda_+-z \\
         1 & 1\\
        \end{array}
\right),
\label{eq:s9}    
\end{equation}
with $\lambda_{\pm}$, the eigenvalues of $T$, are given by
\begin{equation}
\lambda_{\pm}=\frac{1}{2}\left(p+z\pm\sqrt{(p-z)^2+4z}\right).
\label{eq:s10} 
\end{equation}
We perform the calculations for the classical $p$-FASEP in the thermodynamic limit $L\rightarrow\infty$. Correspondingly, the partition function simplifies to $Tr[T^L]\approx\lambda_+^L$. In this limit, it is straightforward to show that the currents for two different modes from Eq.~(\ref{eq:s7}) result in 
\begin{eqnarray}
J_{I}&=& \frac{z^2}{\lambda_+^3 \Delta U}(U_{11}-U_{21})U_{12}, \cr
J_{II}&=& \frac{p z}{\lambda_+^3 \Delta U}(U_{11}-p U_{21})U_{12}, 
\label{eq:s11}
\end{eqnarray}
where $U_{jk}$ correspond to the entry of matrix $U$ [Eq.~(\ref{eq:s9})] in $j$-th row and $k$-th column, $\Delta U$ is the determinant of the matrix $U$. Following few simple intermediate steps, we can arrive at the expressions of the currents given in Eq.~(\ref{eq:jfajin}).  The fugacity $z$ is determined from the density-fugacity relation $\rho=\frac{z}{\lambda_+}\frac{d}{dz}\lambda_+$, as a function of the input parameters $\rho$ and $p$. The corresponding solution $z(\rho,p)$ is provided in Eq.~(\ref{eq:jfajin}) of the main text.

\end{widetext}

\end{document}